\documentclass[
	a4paper, 
	10pt,
	twocolumn
]{article}

\usepackage[
	top=2.5cm, 
	bottom=2.5cm, 
	left=15mm, 
	right=15mm, 
	footskip=1cm, 
	headsep=0.75cm, 
	columnsep=9mm, 
]{geometry}

\usepackage[utf8]{inputenc} 
\usepackage[T1]{fontenc} 

\usepackage[sc]{mathpazo} 
\usepackage{fancyhdr} 
\fancypagestyle{mdoc}{%
 
\fancyhf{} 
\fancyhead[L]{\small\textit{\runninghead}} 
\fancyfoot[C]{\thepage} 
}
\fancypagestyle{plain}{%
 
	\fancyhf{} 
	\fancyfoot[C]{\thepage} 
}

\newcommand{\runninghead}[1]{\renewcommand{\runninghead}{#1}}
\newcommand{\footertext}[1]{\renewcommand{\footertext}{#1}}

\usepackage{titlesec} 
\titleformat
	{\section} 
	[block] 
	{\Large\bfseries\centering} 
	{\arabic{section}.} 
	{0.5em} 
	{} 
	[]
\titleformat
	{\subsection} 
	[block] 
	{\raggedright\large\bfseries} 
	{\arabic{section}.\arabic{subsection}} 
	{0.5em} 
	{} 
	[] 
\titleformat
	{\subsubsection} 
	[runin] 
	{\bfseries} 
	{} 
	{5pt} 
	{} 
	[] 
\titlespacing*{\subsubsection}{0pt}{0.5\baselineskip}{8pt} 
\titlespacing*{\section}{2pt}{0.5\baselineskip}{4pt}

\usepackage{titling} 
\pretitle{\begin{center}\huge\bfseries} 
\posttitle{\end{center}} 
\usepackage{abstract}

\usepackage[backend=biber,
			style=phys,
			biblabel=brackets,
]{biblatex}
\usepackage{graphicx} 
\graphicspath{./} 
\usepackage{caption} 
\usepackage{booktabs} 
\usepackage{enumitem} 
\setlist{noitemsep}

\usepackage[separate-uncertainty=true,
			exponent-product=\cdot,
			list-final-separator={, and },			
			]{siunitx}
\usepackage[version=3]{mhchem}
\usepackage{chemfig}
\usepackage{amsmath}
\usepackage{amsfonts}
\usepackage{amssymb}
\usepackage{upgreek}
\usepackage{mathtools}
\usepackage[final,
			pdfusetitle,
			]
			{hyperref} 
\usepackage[nameinlink]{cleveref}
\makeatletter
\AddToHook{cmd/appendix/before}{\def\cref@section@alias{appendix}\def\cref@subsection@alias{appendix}}
\makeatother

\DeclareFieldFormat{titlecase}{#1}
\runninghead{Bayesian Optimal Design for Non-Linear Parameter Estimation} 

\newcommand{\mva}{\mvec{a}}
\newcommand{\mvu}{\mvec{u}}
\newcommand{\mvx}{\mvec{x}}
\newcommand{\mvy}{\mvec{y}}
\newcommand{\mvp}{\mvec{\varphi}}
\newcommand{\mveps}{\mvec{\varepsilon}}
\newcommand{\mmP}{\mmat{\Phi}}
\newcommand{\mmPp}{\mmP^{\prime}}
\newcommand{\mmPtp}{\mmP^{\prime \mkern-1.5mu\mathsf{T}}}

\newcommand{\tp}{t_+^{\ast}}
\newcommand{\tm}{t_-^{\ast}}
\newcommand{\ts}{t^{\ast}}

\newcommand{\eulernum}{\ensuremath{\mathrm{e}}}

\newcommand{\prob}{\ensuremath{P}}

\newcommand{\condprob}[2]{\prob{}\left( #1 \vert #2 \right)}

\newcommand{\orderof}[1]{\ensuremath{\mathcal{O}\left(#1\right)}}

\newcommand{\fullstop}{\, .}
\newcommand{\comma}{\, ,}

\newcommand{\expect}{\mathbb{E}}
\newcommand{\projector}{\mmat{P}}
\newcommand{\Pperp}{\projector{}_\perp}
\newcommand{\hess}{\mmat{H}}

\newcommand{\mvec}[1]{\vec{#1}}
\newcommand{\mvecnorm}[1]{\left\lVert#1\right\rVert}
\newcommand{\mmat}[1]{\mathbf{#1}}

\newcommand{\idmat}{\mathbb{I}}

\newcommand{\true}[1]{\hat{#1}}
\def\transpose#1 {#1^{\mkern-1.5mu\mathsf{T}}}
\def\hc#1 {#1^{\dagger}}

\DeclareMathOperator{\Tr}{Tr}
\DeclareMathOperator{\Expect}{\mathbb{E}} 
\usepackage{authblk}

\title{Optimal and Adaptive Bayesian Sampling for Non-Linear Parameter Estimation under White Noise}

\makeatletter
\renewcommand{\AB@authnote}[1]{\normalfont#1}

\makeatother
\author[$^{1,}$\thanks{Corresponding author: \href{mailto:lennart.bosch@uni-ulm.de}{lennart.bosch@uni-ulm.de}}]{Lennart H. Bosch}
\author[$^{1,2}$]{Martin B. Plenio}

\affil[1]{Institute of Theoretical Physics, Ulm University, 89081 Ulm, Germany.}
\affil[2]{Center of Integrated Quantum Science and Technology (IQST), Ulm University, 89081 Ulm, Germany.}

\date{
	June 18, 2026
 	}

\renewcommand{\maketitlehookd}{%
	\begin{abstract}
		\noindent The question of optimal experimental design has been addressed in a vast variety of contexts and answered using manifold approaches.
Assuming additive white Gaussian noise, this work applies the Bayesian framework for design optimization to the posterior distribution after marginalization over linear parameters and discusses the implications.
Examples of exponentially decaying signals with and without oscillations complement the discussion.
Application of the examples considered include but are not limited to nuclear magnetic resonance and relaxometry experiments using solid-state spins sensors.

	\end{abstract}
}

\DeclareSIUnit\Molar{M}
\DeclareSIUnit\rpm{rpm}
\usepackage{esint}

\makeatletter
\@for\next:={int,iint,iiint,iiiint,dotsint,oint,oiint,sqint,sqiint,
  ointctrclockwise,ointclockwise,varointclockwise,varointctrclockwise,
  fint,varoiint,landupint,landdownint}\do{%
    \expandafter\edef\csname\next\endcsname{%
      \noexpand\DOTSI
      \expandafter\noexpand\csname\next op\endcsname
      \noexpand\ilimits@
    }%
  }
\makeatother

\begin{document}

\maketitle 


\section{Introduction}
\label{sec:intro}

Non-linear model fitting is a common problem in various fields of science and engineering aiming to estimate unknown parameters from a given set of data samples.
In correspondence with the expected signal model, the exact design of the experiment and sampling scheme may greatly impact the overall achievable precision.
Signal samples acquired at high signal amplitudes relative to the noise floor are intuitively more informative for parameter estimation than samples acquired in signal regions, where the signal is almost completely buried in noise.
For various non-linear models, however, correlations between samples are crucial for high precision and the optimal sampling constitutes a highly non-trivial configuration. 
Systematic approaches to finding the sampling schemes that are expected to yield the highest possible accuracy using the available resources, such as time or number of samples, are collectively referred to as experimental design optimization~\cite{Pukelsheim2006a,Atkinson2007a}.
Which regions of the sampling space exhibit characteristic behavior is typically governed by the true values of underlying model parameters and, consequently, so will the optimal design.
This further motivates the use of adaptive sampling approaches that dynamically update the knowledge of parameter values during the data acquisition and reallocate sampling or further respective resources to improve the estimation precision~\cite{Myung2013a}.

Experimental design optimization has been successfully employed
across a wide range of disciplines, including sensor placement optimization~\cite{Krause2008a}, study design in psychology~\cite{McClelland1997a,Kwon2023a}, and sampling strategy optimization~\cite{Pukelsheim2006a}, where the appropriate optimality criterion fundamentally depends on the research question and parameters of interest.
Over recent years, the techniques have also  increasingly been applied to experiments involving solid-state spin sensors such as the nitrogen-vacancy~(NV) center, featuring applications like magnetometry~\cite{Bonato2016a,Dinani2019a,Dushenko2020a,McMichael2021b}, nuclear spin sensing\cite{Joas2021a}, charge state detection\cite{DAnjou2016a}, and relaxometry~\cite{CaouetteMansour2022a}.
In above mentioned quantum sensor setups, efficient sampling schemes improve the expected precision by focusing on relevant sampling regions, for example, frequency intervals including resonances or time periods before the signal decay by decoherence.

A vast number of optimality criteria are derived from the Fisher information matrix~\cite{Pukelsheim2006a,Atkinson2007a}, which quantifies the information that an observable random variable carries about unknown model parameters.
An alternative class of approaches utilizes the principles of Bayesian parameter estimation where the expected information gain constitutes a natural figure of merit~\cite{Chaloner1995a,Rainforth2024a}.
Within this framework, adaptive schemes emerge naturally through principles of Bayesian updating~\cite{Chaloner1995a,Sebastiani2000a,Myung2013a,Rainforth2024a}.
The computation of the expected information gain, however, typically requires the evaluation of nested integrals, such that practical implementations often rely on numerical approximations or stochastic methods~\cite{Long2013a,Beck2018a,Wu2023a}.

This work considers a signal scenario in which linear signal parameters are eliminated from the estimation procedure through marginalization assuming white Gaussian noise and discusses the consequences for optimal experimental designs derived from the expected information gain.
To this end, we make heavy use of the Laplace approximation to simplify one of the integrals in the expected information gain~\cite{Long2013a,Beck2018a} and perform a series expansion of the remaining expression.
Leading expansion terms yield a very compact expression that facilitates analytical results for exponentially decaying signals and admits compact implementations of adaptive sampling schemes.
Because exponentially decaying signals are a common occurrence in different fields of science, optimal sampling schemes have been discussed in the corresponding literature~\cite{Wigren1991a,Jones1996a,Han2003a,Chardon2010a,Bolzonello2024a}.
The results obtained from the approach presented in this work is capable of reproducing the known results using a more accessible description, complements them with numerical experiments, and expands on the discussions through the adaptive schemes.
Application of above described methods to NV center experiments requires operation under conditions that render the independent white noise a sufficiently accurate model of the noise.
This is achieved either through large photon counting statistics or the use of an ensemble of NV centers~\cite{Dushenko2020a,McMichael2021b,CaouetteMansour2022a}.

\section{Theory}
\subsection*{Model and Analytical Marginalization}
The class of signals considered in this work assumes a model $f(t;\theta, \mvec{a})$ which constitutes a linear superposition with weights $\mvec{a} \in \mathbb{R}^m$ of non-linear basis functions $\varphi_k(t; \theta)$ with $k=1,\dots,m$ that depend explicitly on time and a set of parameters $\theta$.
Taking $n$ samples at discrete points in time $t_j$ with $j=1,\dots,n$, the model compactly reads
\begin{align}
\mvec{f}(\mvec{a}, \theta) = \mmat{\Phi}(\theta) \cdot \mvec{a} \comma \label{eq:signal_model}
\end{align}
where the matrix $\mmat{\Phi} \in \mathbb{R}^{n\times m}$ is composed of column vectors of the non-linear basis functions $\varphi_k(t; \theta)$ evaluated for all $t_j$.
The observed data is modeled as random samples from the likelihood distribution $\prob \big( \mvec{y} \vert \theta, \mvec{a} \big)$, which is conditioned on the signal model through the associated parameters and resembles the nature of the noise.
Further discussions focus on the most widespread noise model of additive white noise for which the likelihood distribution reads
\begin{align}
\condprob{\mvec{y}}{\theta, \mvec{a}} = 
\frac{1}{(2\pi\sigma^2)^{n/2}} \exp \left\lbrace - \frac{1}{2\sigma^2} \mvecnorm{\mvec{y} - \mmat{\Phi} \cdot \mvec{a} }^2 \right\rbrace \fullstop \label{eq:init_likelihood}
\end{align}
Equivalently, the data samples are described as sum of the true signal and background noise as
\begin{align}
\mvec{y} = \true{\mmP} \cdot \true{\mva} + \mvec{\varepsilon} \comma
\label{eq:noisy_samples}
\end{align}
with elements of $\mveps$ being identically independently distributed and hats indicate the use of true parameter values.

For a variety of problems, parameter estimates can efficiently be obtained from the maximum of the likelihood in \cref{eq:init_likelihood} with respect to $\mvec{a}$ and $\theta$~\cite{Kay1993a}.
In contrast, the Bayesian approach aims to derive parameter estimates from the posterior distribution $\condprob{\theta, \mvec{a}}{\mvec{y}}$, instead.
Starting from the noise model, and imposing a prior distribution $\prob (\theta,\mvec{a})$ for the parameters to estimate, the posterior distribution is obtained via Bayes theorem~\cite{Bayes1763a} 
\begin{align}
\condprob{\theta, \mvec{a}}{\mvec{y}} = \frac{\condprob{\mvec{y}}{\theta, \mvec{a}} \cdot \prob (\theta,\mvec{a})}{\prob (\mvec{y})} \fullstop
\end{align}
The prior distribution on the data $\prob (\mvy)$ can merely be understood as normalization constant for the posterior.
Parameter estimates are then often inferred from the mean of the posterior distribution and some measure of the width provides uncertainty estimates
For a detailed discussion of the implications and consequences, we shall refer to the book by Kruschke~\cite[Ch.2]{Kruschke2015a}.

Treating problems of spectral estimation, for example, the linear parameters $\mvec{a}$ are of negligible interest, and following the example of Bretthorst~\cite{Bretthorst1990a,Bretthorst1990b,Bretthorst1990c}, can be marginalized over to obtain a posterior distribution in the non-linear parameters only as
\begin{align}
\condprob{\theta}{\mvec{y}} = \int\limits_{-\infty}^{\infty} \condprob{\theta, \mvec{a}}{\mvec{y}} \mathrm{d}\mvec{a} \fullstop  \label{eq:raw_marg}
\end{align}
To express insignificant prior knowledge of $\mvec{a}$, a natural choice for the prior in $\mva$ is given by a normal prior distribution centered around \num{0}.
Choosing the distribution's width $\tau$ for each element of $\mvec{a}$ sufficiently large such that the prior covers the true values, the prior acts as a gentle regularization without distorting influence of the data. 
Thus, the prior distribution explicitly reads
\begin{align}
\prob (\theta,\mvec{a}) = \frac{1}{(2\pi \tau^2)^{m/2}} \exp \lbrace - \frac{\mvecnorm{\mvec{a}}^2}{2\tau^2} \rbrace \cdot \prob (\theta) \fullstop
\end{align}
Above choice for the prior allows for analytical evaluation of the marginalization in \cref{eq:raw_marg} and, omitting terms that do not explicitly depend on $\theta$, yields
\begin{align}
\begin{split}
-\ln \condprob{\theta}{\mvec{y}} &\varpropto  \frac{1}{2\sigma^2} \transpose\mvec{y} (\idmat - \mmat{\Phi} (\transpose\mmat{\Phi} \mmat{\Phi} + \xi^2 \idmat )^{-1} \transpose\mmat{\Phi} ) \mvec{y} \\
&+ \frac{1}{2} \ln \lbrace \det (\transpose\mmat{\Phi} \mmat{\Phi} + \xi^2 \idmat ) \rbrace - \ln \prob(\theta) \\
&=:F(\theta;\mvec{y}) \comma
\end{split} \label{eq:log_marg_likelihood}
\end{align}
with $\xi:= \sigma/\tau$\footnote{Please note that the introduction of $\xi$ as a plain substitution is only possible if the distribution is explicitly defined as distribution in $\theta$ only and not also in $\tau$ and $\sigma$.}.
The evaluation of \cref{eq:log_marg_likelihood} requires $\transpose\mmP{} \mmP + \xi^2 \idmat$ to be non-singular. 
While large $\xi$ is technically sufficient for regularization, a safe criterion is $n>m$, i.e., there are more data samples than basis functions, and $\mmP $ has full rank, meaning that basis functions are sufficiently orthogonal.
Throughout the rest of this work, constant (full) rank is assumed for $\mmP{} $ for all values of $\theta $ considered, such that local continuity for $F(\theta ;\mvy{} ) $ in $\theta $ is guaranteed.
The expression for the marginal posterior in \cref{eq:log_marg_likelihood} is employed in similar form in sparse classification as well as regression~\cite{Tipping2003a} and recently gained traction in spectral estimation~\cite{Badiu2017a,Hansen2018a}.
In the following section we present systematic expansions of estimators and the covariance which ultimately give rise to an objective function for the optimization of sampling schemes.

\subsection*{Estimator and Covariance Expansion}

\begin{figure*}
\includegraphics[width=\textwidth]{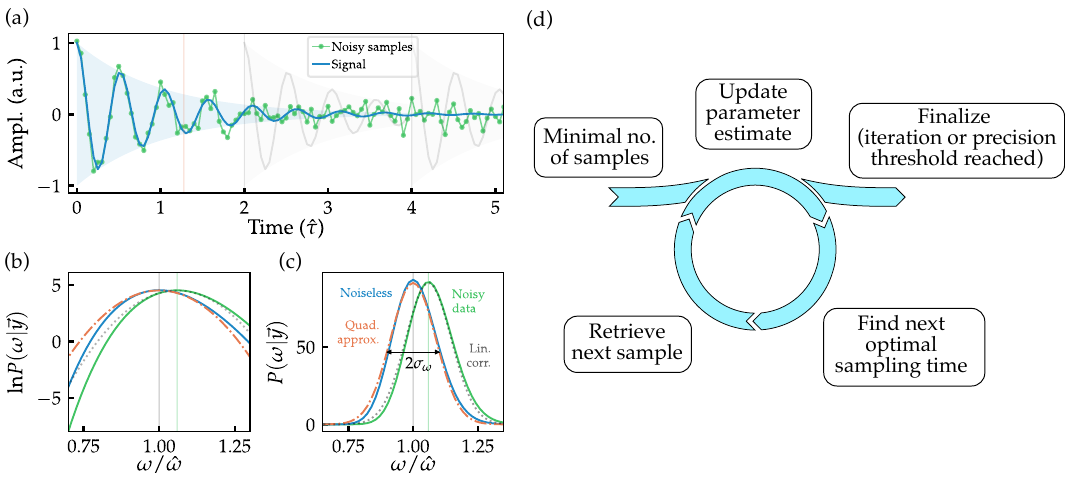}
\caption{
(a) Example plot of an oscillating signal with a single frequency subject to Lorentzian decay over time in units of its lifetime $\true{\tau}$. The gray shaded signal correspond to a re-initialization at two times the lifetime. 
(b) Plot of the posterior distributions logarithm together with its quadratic approximation and the correction by linear expansion of maximum and curvature in the noise. 
(c) Plot of the full posterior distributions shown in (b). The quadratic approximation results in a perfect Gaussian which resembles the original distributions width considerably well. Hats over parameters mark true values. 
(d) Graphical depiction of the adaptive sampling procedure~\cite{Rainforth2024a}.
}
\label{fig:demo}
\end{figure*}

The explicit computation of the posterior mean for generic $\mmP$ is tedious and not expected to yield analytically tractable results. 
Therefore, the remainder of this work follows the example of maximum-likelihood and focuses on the maximum a-posteriori~(MAP) estimator derived from the maximum of $\condprob{\theta}{\mvec{y}}$, instead.
Equivalently, estimators are derived from the minimum $-\log \condprob{\theta}{\mvec{y}}$ denoted by $F(\theta;\mvec{y})$ in \cref{eq:log_marg_likelihood}.
Likewise, this work focuses on uncertainty estimation from the posterior variance instead of quantiles or credibility intervals to allow for a compact analytical treatment.
Because the computation of the moments of the distribution can be cumbersome, nevertheless, upcoming discussions make use of the Laplacian approximation and approximate the variance by the inverse of the distribution's curvature at the maximum.
\Cref{fig:demo}(b) schematically demonstrates the approximation of $F(\theta;\mvec{y})$ up to quadratic order around the maximum and shows the effect on the probability distribution after taking the exponential in \cref{fig:demo}(c).

Above described procedures are straightforward to conduct numerically for a specific realization of the noise $\mvec{\varepsilon}$.
Because $F(\theta;\mvec{y})$ explicitly depends on $\mvec{y}$ and the noise realization, so will the estimate, lifting it to a random variable.
From a theoretical point of view, the estimate is expected to approach a deterministic noiseless estimate of $\theta$ governed by the local minimum of the noiseless objective function $F(\theta;\true{\mmP}\true{\mva})$ upon continuous reduction of the noise variance.
Slowly turning the noise contribution up, again, the estimate will experience a correction that explicitly depends on the exact realization of $\mvec{\varepsilon}$.
To model this behavior theoretically, the estimate is expanded into 
\begin{align}
\theta_\mathrm{MAP} = \theta_0 + \delta\theta \comma
\end{align}
where $\theta_0$ denotes the noiseless estimate and $\delta\theta$ describes the correction explicitly depending on $\mveps$.
Assuming a local optimum, the defining equation for $\theta_0$ and $\delta\theta$ constitutes $\partial_\theta F(\theta_\mathrm{MAP};\mvec{y})=0$.
To separate out the effect of $\delta\theta$, expand the implicit definition around $\theta_0$ up to linear order in $\delta\theta$ as
\begin{align}
0 = \partial_\theta F(\theta_0;\mvec{y}) + \partial_\theta^2 F(\theta_0;\mvec{y}) \cdot \delta\theta + \mathcal{O} \big( \delta\theta^2 \big) \fullstop \label{eq:Pexpansion}
\end{align}
All individual terms of the expansion are required to vanish for the equation to be satisfied. 
Expanding $\delta\theta$ further in powers of the noise as $\delta\theta=\delta\theta^{(1)}+\delta\theta^{(2)}+\dots$ and separating the terms in \cref{eq:Pexpansion} by powers of $\varepsilon$ (indicated by the superscript) yields a system of equations for expansion terms $\delta\theta^{(j)}$ upon setting the corresponding coefficients equal to zero.
For demonstrational purposes, we shall note here that the resulting equations for the deterministic and linear order read
\begin{subequations}
\begin{align}
\partial_\theta F^{(0)}(\theta_0;\mvec{y}) &= 0 \label{eq:defeq1} \\
\partial_\theta F^{(1)}(\theta_0;\mvec{y}) + \partial^2_\theta F^{(0)}(\theta_0;\mvec{y}) \cdot \delta\theta^{(1)} &= 0 \comma \label{eq:defeq2}
\end{align}
\end{subequations}
that now allow to solve \cref{eq:defeq1} for $\theta_0$ and recursively obtain the linear correction $\delta\theta^{(1)}$ from \cref{eq:defeq2}.
Higher orders are obtained from the corresponding extension of above equations.
Ref.~\cite{Yao1994a} employed a similar approach to derive variance estimates of the estimator.
 
Recalling the posterior's logarithm from \cref{eq:noisy_samples} and introducing $\mmat{M} = \idmat - \mmat{\Phi} (\transpose\mmat{\Phi} \mmat{\Phi} + \xi^2 \idmat )^{-1} \transpose\mmat{\Phi} $, the objective function from \cref{eq:log_marg_likelihood} expands into powers of $\mveps$ as
\begin{align}
\begin{split}
F(\theta;\mvec{y}) &= \\ 
\frac{1}{2\sigma^2} &\Big\lbrace 
  \transpose\mveps{} \mmat{M} \mveps
+ \transpose\mveps{} \mmat{M} \true{\mmP} \true{\mva} + \transpose\true{\mva{}} \transpose\true{\mmP} \mmat{M} \mveps 
+ \transpose\true{\mva}{} \transpose\true{\mmP} \mmat{M} \true{\mmP} \true{\mva}
\Big\rbrace \\
&+ \frac{1}{2} \ln \lbrace \det (\transpose\mmat{\Phi} \mmat{\Phi} + \xi^2 \idmat )\rbrace - \ln \prob(\theta) \comma
\end{split}
\end{align}
and one can read off the individual contributions required for \cref{eq:defeq1,eq:defeq2}.
For problems with multiple free parameters compactly described by multidimensional $\theta \in \mathbb{R}^p$, the expansion up to linear order requires the introduction of the gradient and Hessian to end up at the defining equations
\begin{subequations}
\begin{align}
\nabla F^{(0)}(\theta_0;\mvec{y}) &= 0 \label{eq:defeqm1} \\
\nabla F^{(1)}(\theta_0;\mvec{y}) + \hess^{(0)}(\theta_0;\mvec{y}) \cdot \delta\theta^{(1)} &= 0 \label{eq:defeqm2} \comma
\end{align}
\end{subequations}
with elements of the matrix $\hess$ given by $[\hess]_{jk} = \partial_j\partial_k F(\theta_0;\mvec{y})$ and $\partial_j$ denotes the derivative with respect to the $j$-th component of $\theta$.
Higher-order corrections from the expansion in \cref{eq:Pexpansion} require the introduction of higher-order tensors and, therefore, quickly become tedious to evaluate. 

The Hessian used to approximate the posterior distribution and whose inverse provides an estimate of the covariance similarly can be separated into a noiseless expression and corrections depending on the realization of $\mveps$ as 
\begin{align}
\hess (\theta_\mathrm{MAP};\mvec{y}) = \hess_0 + \delta\hess \fullstop
\label{eq:hess_correction}
\end{align}
To identify the contributions, expand the Hessian at the MAP estimator in powers of corrections as
\begin{align}
\hess (\theta_\mathrm{MAP};\mvec{y}) = \hess (\theta_0;\mvec{y}) + \nabla \hess(\theta_0;\mvy) \cdot \delta\theta 
+\dots
\comma
\label{eq:hess_expansion}
\end{align}
where the term $\nabla \hess \cdot \delta\theta$ has to be understood as $\sum_j \partial_j \hess(\theta_0;\mvec{y}) \delta\theta_j$.
Similar as for \cref{eq:defeq1,eq:defeq2}, the expansion terms can be sorted by powers of $\mveps$ to read off the expressions for $\hess_0$ and the contribution to $\delta\hess$ linear in $\mveps$.
For a posterior of the form in \cref{eq:log_marg_likelihood}, the noiseless term is obtained upon setting $\mveps$ in $\mvy$ to zero and plugging in $\theta_0$ for the non-linear parameter.

For the analysis of the scaling behavior of noiseless contribution and corrections, limit the discussion to a scenario under which $\transpose\mmP \mmP + \xi^2 \idmat$ is far from singular around $\theta_0$ ($\mmP$ has full column-rank and $n>m$) and  $\mmP$ is assumed at least twice continuously differentiable in a local environment around $\theta_0$, such that also $F(\theta_0;\mvy)$ is twice continuously differentiable.
In addition, the sampling space is required to be bounded as $t_\mathrm{l} \leq \lbrace t_j \rbrace \leq t_\mathrm{u}$.
Because $\theta_0$ is chosen such that $F(\theta_0;\true{\mmP} \true{\mva})$ exhibits a local maximum, $\hess_0$ is positive definite.
$\hess_0$ is then expected to asymptotically approach the Fisher information and, therefore, grows linearly with $n$.
The correction term is decomposable into a sum of random variables each constructed from elements of $\mveps$ and powers thereof, such that by the central limit theorem, it will exhibit a $\sqrt{n}$ scaling with the number of samples.
Thus, with increasing number of samples, the deterministic contribution increasingly dominates the overall Hessian and becomes an increasingly accurate approximation.
Lifting the constraint on the sampling space to be bounded, one can construct scenarios in which $\hess_0$ achieves super-linear scaling, for example, for the uniform sampling of an oscillating function with constant amplitude~\cite{Bretthorst1988a}.

The scaling behavior of the covariance matrix as inverse of $H(\theta_\mathrm{MAP};\mvy)$ follows from the expansion
\begin{align}
(\hess_0+\delta\hess)^{-1} = \hess_0^{-1} - \hess_0^{-1}\cdot \delta\hess \cdot \hess_0^{-1} + \dots \fullstop \label{eq:cov_expanded}
\end{align}
Recalling the scaling of $\hess_0$ and $\delta\hess$, the leading order term scales as $1/n$ with the next-to-leading order term being suppressed as $1/n^{3/2}$.
Therefore, the leading order term serves as a suitable approximation of the covariance in the case of moderate noise and a reasonable number of samples.

\Cref{fig:demo}(b) schematically shows the transformation of the posterior's logarithm undergoing above described approximation steps with the corresponding probability distribution in \cref{fig:demo}(c).
Starting from the green curve showing the distribution for one particular realization for $\mveps$, the blue curve is obtained by removal of the noise, letting the noiseless estimate from the curve's maximum approximately coincide with the true value.
The expansion up to leading order around the maximum yields the orange dashed curve, 
which upon shifting by the linear correction term $\delta\theta$ results in the gray dotted line.
Even using only the leading order contribution for the Hessian $\hess_0$, the gray dotted line accurately resembles the original distribution.

\subsection*{Optimal and Iterative Sampling}
For a specific realization of the data, the posterior distribution's information content is quantified using the Shannon entropy~\cite{Shannon1948a,Sebastiani2000a} as
\begin{align}
S(\Omega;\mvy ) = - \int \condprob{\theta}{\mvy} \ln \big( \condprob{\theta}{\mvy} \big) \mathrm{d}\theta \fullstop
\end{align}
After invoking the Laplace approximation, the entropy reduces to
\begin{align}
S(\Omega;\mvy ) = - \frac{1}{2} \ln \det(\hess(\theta_\mathrm{MAP}; \mvy)) + \frac{p}{2}(1+\ln(2\pi)) \comma
\label{eq:entropy_gaussian}
\end{align}
where $p$ denotes the dimension of $\theta$. 
Using the separation of terms in the Hessian from \cref{eq:hess_correction} and invoking Fredholm's determinant~\cite{Gohberg2000a} yields
\begin{align}
\begin{aligned}
\ln \det (\hess_0+\delta\hess) &= \ln \det (\hess_0) + \ln \det(1+\hess_0^{-1} \delta\hess ) \\
= \ln \det (\hess_0) &+ \sum\limits_{k=1}^{\infty} \frac{(-1)^{k-1}}{k} \Tr \big( (\hess_0^{-1} \delta\hess )^k \big) \fullstop
\label{eq:Fredholms_det_expansion}
\end{aligned}
\end{align}
Above series expansion exhibits a convergence radius of one and, therefore, puts an upper bound on the allowed value for $\vert \hess_0^{-1} \delta\hess \vert$ for the equality to hold.
A strict condition is provided by \begin{align}
\Vert \delta\hess \Vert \ll \lambda_\mathrm{min}( \hess_0 ) \comma
\label{eq:convergence_condition}
\end{align}
where $\Vert\cdot \Vert$ denotes the operator norm\footnote{Because $\delta\hess$ is assumed to be symmetric, its operator norm is identical to the absolute value of its largest eigenvalue.} and $\lambda_\mathrm{min}$ is the smallest eigenvalue. 
While the scaling of $\hess_0$ and $\delta\hess$ with $n$ discussed in the previous section is generally favorable for the convergence, the suppression of correction terms to the entropy may fail locally if any of the eigenvalues of $\hess_0$ deviates from the expected scaling, directly transpiring to $\det(\hess_0)$.

Assuming \cref{eq:convergence_condition} to hold and feeding the expansion in \cref{eq:Fredholms_det_expansion} back into \cref{eq:entropy_gaussian} allows to compute the expected information entropy as
\begin{align}
\begin{aligned}
\Expect_{\mvy} S(\Omega; \mvy) =  -\frac{1}{2} \ln \det (\hess_0) + &\frac{p}{2}(1+\ln(2\pi)) \\
&+ \mathcal{O}\big(\sigma^2/n\Vert\true{\mva}\Vert^2 \big) \comma
\label{eq:lo_expected_information}
\end{aligned}
\end{align}
For above expansion of $\Expect_{\mvy} S(\Omega; \mvy) $ to yield valid results, a fixed non-linear dimension $p$, fixed number of basis functions $m$, and full column rank of the design matrix $\mmP $ such that $\transpose\mmP{} \mmP $ is far from singular are required.
Similar as for the analysis following \cref{eq:hess_correction,eq:hess_expansion}, under additional assumptions about $\theta_0$ being an identifiable parameter value and bounded or otherwise controlled leverage of the sampling design, the leading order hessian obeys 
\begin{align*}
\hess_0 = \mathcal{O} \Big( \frac{n \Vert \true{\mva} \Vert^2}{\sigma^2} \Big)
\end{align*}
and the leading order zero-mean noise correction follows
\begin{align*}
\delta\hess = \mathcal{O}_P \Big( \frac{\sqrt{n} \Vert \true{\mva} \Vert}{\sigma} \Big) \fullstop
\end{align*}
The expectation value then eliminates contributions linear in the noise the lowest-order surviving term in the combination of both of the above scalings in $\hess_0^{-1}\delta\hess $ finally exhibits the scaling indicated in \cref{eq:lo_expected_information}. 
As rule of thumb, the approximation by the noiseless hessian term is accurate either for a small number of samples as long as either the SNR is significantly larger than \num{1} or $n$ is sufficiently large to compensate for the lack of SNR.

In addition to the statistical corrections, the expression for the expected information entropy in \cref{eq:lo_expected_information} deviates from the true value due to the Laplace approximation.
The scaling behavior of the deviation can be obtained by comparison of higher order terms in the expansion of the posterior's logarithm and is expected to scale linearly in $\sigma/\Vert\true{\mva}\Vert\sqrt{n}$~\cite{Katsevich2024a}.
While this error may dominate over the statistical corrections in some scenarios, the total error is still expected to be suppressed polynomially with increasing number of samples.

The optimization of the sampling scheme, or more generally the experimental design aims to maximize the information entropy which may be approximated by the leading-order term in \cref{eq:lo_expected_information}. 
The monotonicity of the logarithm allows to further strip the expression to the maximization problem
\begin{align}
\max_{ \lbrace t_j \rbrace \in [t_a,t_b]} \det \hess_0 \fullstop \label{eq:maxdetH0}
\end{align}
Above expression weighs each sample equally and, thus, optimizes the sampling scheme for a predefined number of samples.

For selected scenarios there exists an analytic optimum while for the general case one has to resort to numerical optimization, typically including some parameter bounds or further constraints on relative distances between sampling times.
The extension of the objective function to scenarios of varying time-cost per sample requires the distinction between destructive measurements and non-destructive measurements.
The first class requires reinitialization of the system after each sample and the time cost of the experiment corresponds to the sum of each individual sampling times plus the time for read-out and initialization.
The independence of these experiments typically guarantees statistical independence of the noise terms even for samples taken at identical time difference to the respective system initialization.
The second class allows to sample at contiguous time points from the same signal and the time-cost amounts to the time required until the last sample is recorded.
Here the sampling rate is typically limited by the measurement device which may exceed the noise bandwidth and the assumption about statistical independence could break down.
In example A of \cref{sec:results}, we explore a modification of the objective function to account for repetition of identical non-destructive uniform sampling.
A generalization of the objective function as stated in \cref{eq:maxdetH0} to weigh sampling cost individually, however, requires a fundamentally different approach to quantification of the precision and goes beyond the scope of this work.
Since the expression for the information entropy features involved maps that combine information from different sampling times, reasonable heuristic regularization terms to complement the objective function in \cref{eq:maxdetH0} are highly non-trivial.
The similarity of the problem stated in \cref{eq:maxdetH0} to the D-optimality criterion in the literature on optimal experimental design~\cite{Han2003a} is no coincidence but a consequence of $\hess_0$ approaching the Fisher information matrix in the limit of dominant likelihood (vast amount of data or uninformative prior) and, thus, solutions of the corresponding optimization problem are expected to align.

For all data following the relation in \cref{eq:signal_model}, the value of $\det(\hess_0)$ implicitly depends on the true value of the underlying signal parameters through its dependence on $\theta_0$.
This dependence can be explained intuitively using an example of a decaying signal with stationary noise floor.
The optimal sampling scheme requires knowledge of the signal life time to avoid regions after complete decay of the signal as these samples evidently carry no usable information.
While this dependence is easily handled in a theoretical study, it resembles an issue for the practical implementation of any optimized sampling scheme if little to no knowledge about the true parameters is available.
The problem gives rise to adaptive design optimization methods as reviewed in~\cite{Myung2013a}, for which an experiment starts with a rough estimate of the parameters or even with random sampling and the parameter estimate and the experimental configuration is adjusted once data has arrived.
\Cref{fig:demo}(d) schematically shows a formalization of this concept in which one or more samples are recorded alternatingly with updates of the parameter estimates and adjusting the sampling strategy until desired precision or a sampling threshold is satisfied.

\begin{figure}
\centering
\includegraphics[scale=1]{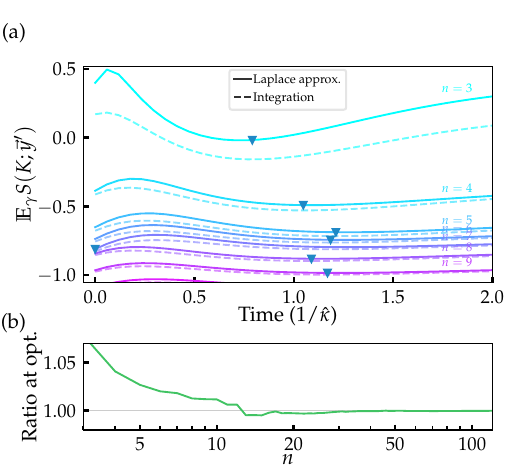}
\caption{
(a) Expected information entropy over the latest sampling time as obtained via nested integration and the leading-order term of the Laplace approximation while keeping all previous fixed for the simple exponential model discussed in Example A of \cref{sec:results}. The next sample is taken each at location of the inverted triangles. 
(b) Ratio of the expected information entropy from Laplace approximation and integration at the curves optimum, i.e., at the inverted triangles.}
\label{fig:inf_ent_iteration}
\end{figure}

Focusing on the case of one sample at a time, the ideal strategy is to choose the sampling time such that it has the highest positive impact on entropy taking into account all previous samples.
The arrival of a new sample with value $\gamma$ is described by the transformation of design matrix and data vector as
\begin{align}
\mmP \to \mmP^\prime = \begin{pmatrix} \mmP \\ \transpose\mvp{} \end{pmatrix} \comma \ \text{and} \
\mvy \to \mvy^{\prime} = \begin{pmatrix} \mvy \\ \gamma \end{pmatrix} \comma
\label{eq:mvy_update}
\end{align}
where $\mvec{\varphi}$ describes the basis functions evaluated at the next sampling time.
Inserting the expansion from \cref{eq:Fredholms_det_expansion} into \cref{eq:entropy_gaussian}, the change in information entropy reads~\cite{Rainforth2024a}
\begin{align}
\begin{aligned}
S(\Omega;\mvy) &- S(\Omega;\mvy^{\prime}) = - \frac{1}{2} \big( \ln \det(\hess_0)- \ln \det(\hess_0^{\prime}) \big) \\
& -\sum\limits_{k=1}^{\infty} \frac{(-1)^{k-1}}{2k} \Tr \big(  (\hess_0^{-1} \delta\hess )^k - (\hess_0^{\prime-1} \delta\hess^\prime )^k  \big) 
\end{aligned}
\end{align}
Consequently, leading order terms of the expected information gain upon adding a sample come down to
\begin{align}
\begin{aligned}
\expect_{\gamma} \big[ S(\Omega;\mvy) & - S(\Omega;\mvy^{\prime})  \big] \\
&= \frac{1}{2} \ln\Big( \frac{\det(\hess_0^{\prime})}{\det(\hess_0)}\Big) + \mathcal{O}_P\big(\sigma/\sqrt{n}\Vert\true{\mva}\Vert \big) \comma
\end{aligned}
\end{align}
where $\expect_\gamma$ denotes the expectation value over the newly acquired sample and $\mathcal{O}_P$ describes the scaling of statistical correction terms originating from previous samples.
Corrections from second moments coming from $\expect\gamma$ scale as $\sigma^2/n \Vert \true{\mva} \Vert^2$ and are, therefore, not explicitly mentioned here.
Iteratively finding the next sample based on the greatest benefit to the estimation given all previous samples then comes down to maximization of $\det(\hess_0^{\prime})$ on some interval $[t_a, t_b]$, where basis functions in $\hess_0^{\prime}$ are evaluated using the current best parameter estimates for $\theta$.
For practical purposes often $t_a=0$ and $t_b<\infty$ and the optimum for various scenarios can very well be located on the interval boundary.
The use of the best parameter estimates in place of $\theta_0$ technically results in the evaluation of $\hess_0+\delta\hess$ instead of $\hess_0$ and, therefore, accounts for parts of the random corrections.
For demonstration purposes, \cref{fig:inf_ent_iteration}(a) exemplary depicts the expected information upon arrival of a new sample as a function of the sampling time during each step of the adaptive procedure for a simple exponential decay signal treated also in example A of \cref{sec:results}.
In addition to the expected information entropy computed from the Laplace approximation, \cref{fig:inf_ent_iteration}(a) also features the expected information entropy computed via numerical integration with the ratio of solid and dashed curves evaluated at corresponding optima in \cref{fig:inf_ent_iteration}(b).

For statistically independent samples, the noiseless Hessian typically separates into contributions from different samples and if the change constitutes a rank-1 update of the form
\begin{align}
\hess_0^{\prime} = \hess_0 + \mvec{v} \cdot \transpose\mvec{v} \comma \label{eq:hess_rank1}
\end{align}
the change in entropy simplifies to~\cite[Sec.6.2]{Meyer2000a}
\begin{align}
\begin{aligned}
\expect_{\gamma} \big[ S(\Omega;\mvy) &- S(\Omega;\mvy^{\prime}) \big] \\
&= \frac{1}{2} \ln \big( 1 + \transpose\mvec{v} \hess_0^{-1} \mvec{v} \big) + \mathcal{O}_P \big(\sigma/n\Vert\true{\mva}\Vert \big) \fullstop \label{eq:entropy_diff}
\end{aligned}
\end{align}
When the leading order Hessian is used as approximation to compute the estimation precision, finding the next sample comes down to a maximization of the leading term in \cref{eq:entropy_diff} with respect to the new sampling times, which solely appears in $\mvec{v}$ through $\mvec{\varphi}$.
The non-linear dependence of $\mvec{v}$ on the next sampling time, however, leaves no options but standard numerical optimization routines.\footnote{Knowledge of the eigenvector of $\hess_0^{-1}$ to the largest eigenvalue permits the reformulation as non-linear least-squares problem, but this eigenvector is generally also unknown, requiring extra computational effort before the actual optimization.}

The impact of an additional data sample on the posterior's maximum is typically smaller than the posterior's current width and decreases with the total number of samples.
In a practical setting, the parameter update is easiest computed using a trust region-based batch optimization routine with the previous estimator as starting point.
Because the numerical optimizer is called repeatedly on barely changing data, the convergence criteria can be chosen rather loosely to reduce the number of iterations per parameter update and speed up the overall procedure.
As demonstrated in \cref{app:asym_post_update}, the independent statistic nature of samples admits the logarithm of the posterior distribution in the asymptotic regime to separate as
\begin{align}
F(\theta;\mvy^{\prime}) = F(\theta;\mvy) + \Delta(\theta;\mvy^{\prime}) \comma \label{eq:F_separate}
\end{align}
with $\Delta$ denoting the contribution imposed by the additional sample.
Assuming the optimizer $\theta_0$ before the update known such that $\partial_j F(\theta_0;\mvy)=0$, the (Quasi-)Newton update step reduces to
\begin{align}
\theta_0 \to \theta_0 - \big(\hess^{\prime}(\theta_0)\big)^{-1} \cdot \nabla \Delta(\theta_0;\mvy^{\prime}) \fullstop
\end{align}
The expression for $\hess^{\prime}$ can here easily be approximated using the leading order hessian $\hess_0^{\prime}$ after updating the best guess of $\mva$ used to approximate the true values taking into account the newly acquired sample.
While above expression may be interesting for an analytical investigation, the well-known Levenberg-Marquardt~\cite{More1978a} method to compute the update step exhibits computational advantages because it does not require explicit computation of the inverse but relies on evaluation of the pseudoinverse of the Jacobian matrix of $\Pperp^{\prime} \mvy^{\prime}$ evaluated at $\theta_0$, only.


While a change in the sampling strategy may influence the rate of convergence of the estimator with increasing number of samples, the asymptotic bias remains unaffected.
Because the iterative approach may also distribute samples uniformly it is capable of matching the precision of uniform sampling.
There are some scenarios in which it is apparent, that above presented greedy algorithm for increasing number of samples asymptotically approaches the global optimal value with respect to all sampling times, e.g., in example A of \cref{sec:results}.
In the general case, however, greedy optimization may fail due to the same reasons as gradient-based optimizers, such as local optima, aliasing, and poor initial estimates.
\Cref{app:num_stab} provides insights into the convergence behavior of parameter estimates under adaptive sampling for the examples discussed in \cref{sec:results}.

\subsection*{Asymptotic Limit Posterior}
The marginal distribution in \cref{eq:log_marg_likelihood} represents an involved expression for a wide variety of basis functions and prior distributions on $\theta$.
Some simple general statements are possible, nevertheless, when the posterior is treated in the context of what is henceforth referred to as the asymptotic limit.
The asymptotic limit describes the case of large signal to noise ratio $\Vert  \true{\mva} \Vert / \sigma$ or a large number of data samples or both.
While these assumptions above sound like a stringent criterion, they typically are satisfied whenever a human observer can identify a signal pattern in noisy data with the bare eye.\footnote{Being able to see a signal or pattern in the data is typically a minimal requirement to convince researchers of the existence of a signal. 
Performing parameter estimation using a signal whose existence is uncertain in most cases yields unsatisfying results and we do our best to avoid this regime.}
Under these conditions, it is also sufficient to assume an uninformative prior on both, $\mvec{a}$ and $\theta$.
The expression for the posterior is then well approximated by dropping terms independent of $\mvy$ in \cref{eq:log_marg_likelihood} and letting $\xi\to 0$ to find
\begin{align}
-\ln \condprob{\theta}{\mvec{y}} \approx  \frac{1}{2\sigma^2} \transpose\mvec{y} (\idmat - \mmat{\Phi} (\transpose\mmat{\Phi} \mmat{\Phi} )^{-1} \transpose\mmat{\Phi} ) \mvec{y} \fullstop
\end{align}
This expression can be compactly rewritten by noting that $\projector(\theta)=\mmat{\Phi} (\transpose\mmat{\Phi} \mmat{\Phi} )^{-1} \transpose\mmat{\Phi} $ describes the projector onto the range of $\mmat{\Phi}$ and introducing the orthogonal projection $\projector_\perp = \idmat - \projector$ such that
\begin{align}
-\ln \condprob{\theta}{\mvec{y}} \approx \frac{1}{2\sigma^2} \Vert \projector_\perp \mvec{y} \Vert^2 =: F_\mathrm{as} (\theta;\mvec{y}) \label{eq:Fas} \fullstop
\end{align}
Above expression is well-known as the Variable Projection objective function~\cite{Golub1973a,Kaufman1975a,OLeary2013a} which is obtained upon solving the LS problem to estimate $\mvec{a}$ assuming fixed $\theta$ and plugging the result back into the objective function for the non-linear LS problem.
Its emergence from marginalization of a normal distribution has been noted in the literature and is not trivial but a consequence of the Gaussian structure of the posterior with respect to the linear parameters~\cite{Bosch2026a}.

For further insights into the estimation behavior of $F_\mathrm{as}$, the expression for $F_\mathrm{as}$ from \cref{eq:Fas} is used to evaluate the defining equation \cref{eq:defeqm1} to find
\begin{align}
0 \stackrel{!}{=} \partial_j F_\mathrm{as}^{(0)} (\theta_0) = \frac{2}{2\sigma^2} \transpose(\Pperp(\theta_0)\true{\mmP}\true{\mva}) (\partial_j \Pperp(\theta_0) \true{\mmP} \true{\mva})  \fullstop
\end{align}
The projection $\Pperp(\theta_0)\true{\mmP}$ trivially yields zero if $\Pperp$ is constructed using $\mmP(\theta_0)=\true{\mmP}$.
Therefore, above equation is satisfied if $\theta_0$ is chosen at the true value $\true{\theta}$ and the estimator proves consistent to leading order under identifiability and regularity assumptions and asymptotically approaches the true value.
Matrix elements of the leading order Hessian $\hess_0$ for $F_\mathrm{as}^{(0)}(\true{\theta})$ required to compute the correction to the estimator as in \cref{eq:defeqm2} and the leading order covariance matrix in \cref{eq:cov_expanded} reduce as
\begin{align}
\begin{aligned}
\partial_k \partial_j F_\mathrm{as}^{(0)} (\true{\theta}) &= \frac{1}{\sigma^2} \partial_k \big[ \transpose(\Pperp(\true{\theta})\true{\mmP}\true{\mva}) (\partial_j \Pperp(\true{\theta}) \true{\mmP} \true{\mva}) \big]\\
\begin{split}
&=\frac{1}{\sigma^2} \transpose(\partial_k\Pperp(\true{\theta})\true{\mmP}\true{\mva}) (\partial_j \Pperp(\true{\theta}) \true{\mmP} \true{\mva}) \\
&\qquad +  \frac{1}{\sigma^2} \transpose(\underbrace{\Pperp(\true{\theta})\true{\mmP}}_{=0}\true{\mva}) (\partial_k\partial_j \Pperp(\true{\theta}) \true{\mmP} \true{\mva})
\end{split} \\
&= \frac{1}{\sigma^2} \transpose(\partial_k\Pperp(\true{\theta})\true{\mmP}\true{\mva}) (\partial_j \Pperp(\true{\theta}) \true{\mmP} \true{\mva}) \fullstop
\end{aligned}
\end{align}
Rewriting the derivative of $\Pperp$~\cite{Golub1973a} evaluated for the true parameters simplifies above expression further to
\begin{align}
\begin{split}
\partial_j \Pperp (\true{\theta}) \true{\mmP} \mva &= -\big[ \Pperp \partial_j \mmP(\true{\theta}) {\true{\mmP}^\dagger \true{\mmP}} 
+ {\transpose(\Pperp\partial_{j}\mmP(\true{\theta})\true{\mmP}^{\dagger}) \true{\mmP}} \big] \true{\mva} \\
&= -\Pperp \partial_j \mmP  \true{\mva} \ \Big\rvert_{\theta=\true{\theta}}
\fullstop
\end{split} \label{eq:der_Pperp_simplified}
\end{align}
The final expression then compactly reads
\begin{align}
\hess_0 = \mmat{L} \cdot \transpose\mmat{L} \label{eq:leadingorder_hessian}
\end{align}
with $\transpose\mmat{L} = \frac{1}{\sigma} \big( \Pperp \partial_1\true{\mmP} \true{\mva}, \  \Pperp \partial_2 \true{\mmP} \true{\mva}, \dots \big)$.

To gain some understanding of the nature of above Hessian, consider a scenario of one-dimensional $\theta$ and a single basis function $\mvec{\varphi}$.
The only element of the Hessian matrix then reduces to 
\begin{align}
h_1 = \frac{\true{a}^2}{\sigma^2} \Vert \Pperp \partial_j \mvec{\varphi}(\true{\theta}) \Vert^2 \fullstop \label{eq:hess_single_par01}
\end{align}
The vector whose norm squared determines the precision explicitly reads
\begin{align}
 \Pperp \partial_j \mvec{\varphi}(\true{\theta}) = \Big[ \partial_j \mvec{\varphi} - \frac{\transpose\mvec{\varphi} \partial_j\mvec{\varphi}}{\transpose\mvec{\varphi} \mvec{\varphi}} \mvec{\varphi}\Big]_{\theta=\true{\theta}} \label{eq:hess_single_par02}
\end{align}
and describes the component of $\partial_j \mvec{\varphi}$ orthogonal to $\mvec{\varphi}$ as, e.g., obtained from the Gram-Schmidt procedure.
In conclusion, the precision is governed by how much the derivative diverts the basis function from the basis function itself.
Generalized to multiple basis functions, the precision is given by the projection of the weighted derivative $\partial_j \mmP(\true{\theta}) \true{\mva}$ onto the orthogonal range of $\mmP$.

The evaluation of the leading order expression for the Hessian in \cref{eq:leadingorder_hessian} requires knowledge of the true parameter values for $\theta$ and $\mva$, admitting theoretical studies of the optimal sampling approach.
With increasing amount of data, however, estimates of $\theta$ and $\mva$ derived from numerical optimization of $F_\mathrm{as}$ and the linear LS problem, respectively, are sufficiently accurate such that $\hess_0$ evaluated with the estimates yields an accurate approximation of the Hessian for uncertainty estimation.
Substitution of the ordinary LS estimates for $\mva$ from $\mmP^{\dagger}\mvy$ recovers the well-known approximation to the Hessian matrix used in the Levenberg-Marquardt algorithm~\cite{More1978a}.

As demonstrated in \Cref{app:asym_post_update}, $F_\mathrm{as}$ separates as in \cref{eq:F_separate} and updates to the leading order Hessian constitute a rank-1 update as in \cref{eq:hess_rank1} with elements of the vector $\mvec{v}$ given by
\begin{align}
\big[\mvec{v}\big]_j = \frac{1}{\sigma\sqrt{1+\transpose\mvu{} \mvu}} 
\begin{pmatrix}
\mvu \\ -1
\end{pmatrix}^{\intercal}
\cdot 
\mvx_j^{\prime} \comma
\end{align}
where $\mvec{u} = \transpose(\mmP^\dagger) \mvp$ and $\mvx_j^{\prime} = \partial_j \mmP^{\prime}(\true{\theta}) \true{\mva}$.
The full matrix element of the leading order Hessian then reads
\begin{align}
\big[\hess_0^{\prime}\big]_{ij} = \big[\hess_0\big]_{ij} + \frac{(\transpose\mvu{} \mvx_i - \eta_i)(\transpose\mvu{} \mvx_j - \eta_j)}{ \sigma^2(\transpose\mvec{u} \mvec{u} + 1)} \comma
\label{eq:hess_update}
\end{align}
where $\eta_j = \transpose (\partial_j\mvec{\varphi}) \true{\mva}$.
While it is possible to isolate the contribution from the previous samples, some of the information is still explicitly contained in the update through $\mmP$.
This behavior explicitly distinguishes the procedure described here from Bayesian Updating for conjugate priors, where information in accumulated representation combined with the new data suffices to fully define the posterior after the update.

\section{Examples}
\label{sec:results}

\subsection*{Example A: Simple Exponential Decay} \label{ex:expdecay}
The first example in this work addresses decay rate estimation for signals described by a simple exponential decay, which emerges in various fields of physics~\cite{Han2003a}. 
In spin physics, the exponential decay accurately describes longitudinal relaxation effects utilized, for example, in relaxometry experiments~\cite{Gong2026a}.
Let $\kappa$ denote the decay rate such that observed signal samples are described by
\begin{align}
y_t = \true a \eulernum^{-\true{\kappa} t} + \varepsilon_t \fullstop
\end{align}
Therefore, the estimation problem targets only a single unknown parameter and is compactly described by a single basis function.
To derive effective sampling strategies, compute the elements of the leading order Hessian by expanding the projector in \cref{eq:leadingorder_hessian} and simplifying the terms to
\begin{align}
\begin{aligned}
\big[ H_0\big]_{jk} = \frac{1}{\sigma^2} &\big\lbrace  \transpose(\partial_k\true{\mmP}\true{\mva}){} \partial_j \true{\mmP} \true{\mva} \\
&- \transpose{( \transpose{\true{\mmP}} \partial_k \true{\mmP} \true{\mva})} (\transpose\true{\mmP}{}  \true{\mmP})^{-1} ( \transpose{\true{\mmP}} \partial_j \true{\mmP} \true{\mva}) \big\rbrace \fullstop \label{eq:Hdiag_simplified}
\end{aligned}
\end{align}
Evaluated at the true value for $\kappa$, the individual terms for the only basis functions read
\begin{subequations}
\begin{align}
\Vert \partial_\kappa \true{\mmP} \true{\mva} \Vert^2 &= \true{a}^2 \sum\limits_{j=1}^n t_j^2 \eulernum^{-2\true{\kappa}t_j} \\
(\transpose\true{\mmP}{}  \true{\mmP})^{-1} &= \Big( \sum\limits_{j=1}^n \eulernum^{-2\true{\kappa}t_j} \Big)^{-1} \\
\Vert \transpose{\true{\mmP}} \partial_\kappa \true{\mmP} \true{\mva} \Vert^2 &= \true{a}^2 \Big( \sum\limits_{j=1}^n -t_j \eulernum^{-2\true{\kappa}t_j} \Big)^2 \fullstop
\end{align}
\end{subequations}
Upon insertion into \cref{eq:Hdiag_simplified}, the only component of the Hessian simplifies to
\begin{align}
h_\kappa = \frac{\true{a}^2}{\sigma^2} \big[ S_2 - \frac{S_1^2}{S_0} \big] \ \ \text{with} \ \
S_k = \sum\limits_{j=1}^n t_j^k \eulernum^{-2\true{\kappa} t_j} \fullstop \label{eq:expmodel_general}
\end{align}
Factoring out $S_0$, the term in parenthesis resembles a variance term in $t$ with weights $\eulernum^{-2\true{\kappa}t}/S_0$, which implies that its value is expected to increase with increasing spread in the discrete sampling points $t_j$. 
On the other hand, weights and prefactor $S_0$ exponentially decay with increasing $t_j$, such that there must exist a sweet spot for the choice of $t_j < \infty$.

\subsubsection*{Uniform Sampling}
To evaluate \cref{eq:expmodel_general} for equidistant sampling, reparametrize the problem by using $t_j = j\cdot \Delta t $ and sum over $j=0,\dots,n-1$ to find
\begin{align}
h_\kappa = (\Delta t)^{2}\frac{\true a ^2}{2\sigma^2} \frac{\eulernum^{\beta }}{\sinh^3(\beta )} \frac{\sinh^2(n\beta ) - n^2 \sinh^2(\beta )}{\eulernum^{2n\beta } - 1} \comma \label{eq:expmodel_equi}
\end{align}
with $\beta$ denoting $\Delta t\true\kappa$.
The associated uncertainty is plotted for fixed $\Delta t$ and $\true{\kappa}=1$ over the number of samples in \cref{fig:exp_iterative}(a) together with numerical estimates of the estimated uncertainty using synthetic data.
To find the sampling interval that achieves maximal precision for a predefined number of samples, above \cref{eq:expmodel_equi} has to be optimized with respect to $\Delta t$. 
To simplify the treatment, one can equivalently multiply $h_\kappa$ with by $\true{\kappa}^2$ and optimize with respect to $\beta$.
Due to the transcendental nature of \cref{eq:expmodel_equi}, however, the maximum can only be expressed as expansion in $n$ yielding
\begin{align}
\Delta t_\mathrm{opt} \true\kappa{} = \num{2.0175} \frac{1}{n} + \num{2.5705} \frac{1}{n^2} + \num{0.4594} \frac{1}{n^3} + \orderof{\frac{1}{n^4}} \fullstop \label{eq:dt_expansion}
\end{align}
Thus, for optimal precision from a single run with a predefined number of equidistant samples, the sampling interval should be chosen such that the last samples resides approximately at two times the signal lifetime.
A plot of the associated uncertainty normalized to its minimum is provided in \cref{fig:exp_iterative}(b), clearly indicating its minimum close to $2/\true{\kappa}$ and the potential relative increase in uncertainty. 

For experimental setups that rely on signal averaging over multiple runs to improve the estimation quality, this raises the question:
If we are given a total time $T$ and within this time frame, the experiment can be stopped and restarted arbitrarily often with the initial amplitude, what is the best time to reinitialize?
$k$ repetitions of the sampling at the same times effectively impacts the posterior distribution by transformation of the noise parameter as $\sigma \to \sigma/\sqrt{k}$.
Denoting $T$ the total time available, one can perform $k = T/(\Delta t n)$ repetitions with sampling interval $\Delta t$ and $n$ samples each.
To find the optimal time spent per repetition, substitute $\sigma^2 \to \sigma^2 \Delta t n / T$ in $h_\kappa$ from \cref{eq:expmodel_equi} accordingly and optimize with respect to $n$.
While $n$, technically, can only take integer values, the expression in \cref{eq:expmodel_equi} extends for continuous $n$ and an expansion of the optimizer in $\beta$ results in
\begin{align}
n_{\mathrm{opt}} \approx \num{2.0175} \beta^{-1} + \num{0.5624} \beta - \num{0.1449} \beta^{3} + \orderof{\beta^5} \fullstop \label{eq:nopt}
\end{align}
The leading order coefficient is identical to the coefficient in \cref{eq:dt_expansion} and we conclude, likewise, that the optimal time spent per repetition is approximately given by two times the signal lifetime.
When choosing the time spent per iteration is offset from the optimal value, the overall reduction in uncertainty approximately behaves as the curve shown in \cref{fig:exp_iterative}(b).

\subsubsection*{Non-Uniform Sampling}
Proceeding to a more general case, by the form of \cref{eq:expmodel_general} we expect to observe a higher precision if $n$ samples can be freely distributed in $[0,\infty)$.
Because the expression in \cref{eq:expmodel_general} is invariant under permutation of samples, the global optimizer is highly degenerate. 
While at $t=0$ the signal amplitude is the largest on the feasible interval, taking all samples on the boundary space yields $h_\kappa=0$ and, thus, can be safely excluded from the set of possible solutions.
As discussed in \cref{app:optsamp_expdecay}, the optimal configuration for fixed $n$ is composed of samples taken at either of the points $t=0$ and $t=t^{\ast}$, only, where $t^{\ast}$ is implicitly defined through
\begin{align}
t^{\ast} = \frac{1}{\true{\kappa}} + \frac{S_1}{S_0} \fullstop
\end{align}
To find the recipe describing the maximizer, one now can reparametrize the problem in terms of finding the number $m$ of $t_j$ assuming the value $0$ (with remaining $n-m$ samples taking $t^{\ast}$) that admits the maximual value of $h_\kappa$.

For finite $n$, the results directly depend on the exact value of $n$ and may not admit a closed form solution.
In the asymptotic limit of large $n$, such that $r=m/n$ can be treated approximately as continuous variable in $(0,1)$, the $S_l$ may be rewritten as
\begin{subequations}
\begin{align}
S_0 &= n r + n(1-r) \eulernum^{-2\true{\kappa} \ts } \\
S_1 &= n(1-r) \ts \eulernum^{-2\true{\kappa} \ts } \\
S_2 &= n(1-r) (\ts)^2 \eulernum^{-2\true{\kappa} \ts } \fullstop
\end{align}
\end{subequations}
Inserting these expressions back into \cref{eq:expmodel_general} yields
\begin{align}
h_\kappa = \frac{\true{a}^2}{\sigma^2} \frac{n}{\true{\kappa}^2} \left( \frac{(\true{\kappa} \ts)^2 (1-r)r}{1+r(\eulernum^{2\true{\kappa} \ts}-1) } \right) \fullstop \label{eq:exp_optimal_precision}
\end{align}
The optimum of above expression with respect to $\ts >0$ and $r\in (0,1)$ is achieved by finding the only critical value for $r$ from the first derivative and plugging the result back in to solve for $\ts$.
This procedure admits the optimizer
\begin{align}
\begin{aligned}
\ts_\mathrm{opt} &= \big(1+W_0(\eulernum^{-1})\big) /\true{\kappa} \approx \num{1.2785}/\true{\kappa} \comma \\
r_\mathrm{opt} &= \frac{1}{1+\eulernum^{\true{\kappa}\ts_\mathrm{opt}}}  \approx  \num{0.2178} \comma
\end{aligned}
\end{align}
where $W_0$ denotes the principal branch of the Lambert W-function.

In conclusion, the optimal precision $h_\kappa$ is achieved upon collapsing all $t_j$ into the points approximately at $\lbrace 0, 1.2785/\true{\kappa} \rbrace$ with relative weights of $r=\qty{21.78}{\percent}$ and $1-r$, respectively.
In practical terms, this recipe proposes to take one in five samples at $t=0$ and the remaining at approximately \num{1.28} the signal lifetime.
For certain experiments, such hyperpolarized NMR, different relative weights can also be achieved by appropriate choice of the flipping angles.
Compared to equidistant sampling with the same number of samples $n$ such that the last sample resides at $t=2/\true{\kappa}$, the uncertainty is reduced by approximately \qty{25.8}{\percent}, and compared to taking half the points at $t=0$ and the remaining at $t=1/\true{\kappa}$ by about \qty{12.4}{\percent}.
Because of the equivalence between minimal estimation uncertainty and maximal Fisher information, above result can identically be retrieved for a c-optimal design deduced from the Fisher information matrix~\cite{Jones1996a,Han2003a}.

\subsubsection*{Iterative Sampling}
\begin{figure*}
\includegraphics[width=\textwidth]{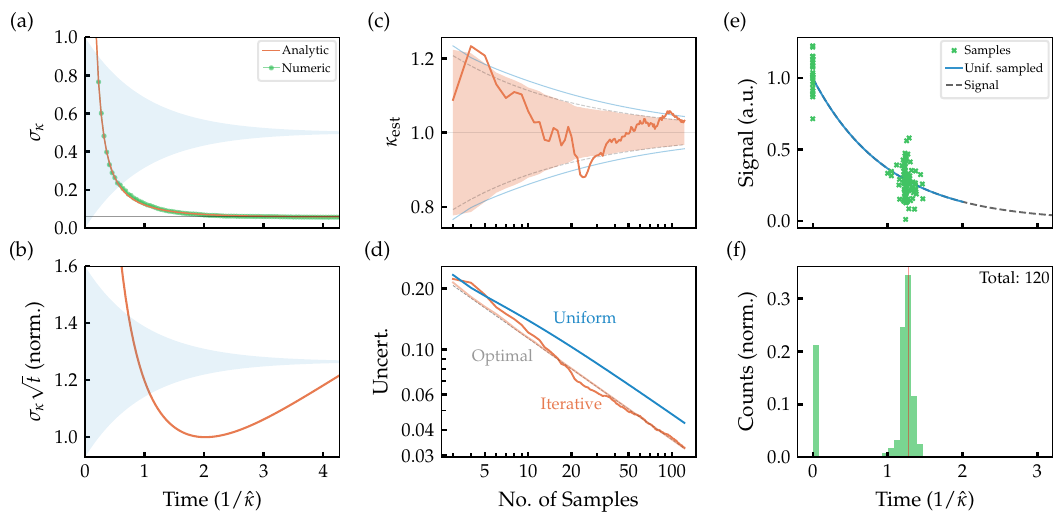}
\caption{
(a) Analytical uncertainty from \cref{eq:expmodel_general} and as obtained by numerical fit of synthetic data with artifical noise for $\true{\kappa} = \num{1}$. The shaded area is drawn to indicate the remaining signal amplitude. 
(b) Analytic uncertainty multiplied by $\sqrt{t}$ (normalized) to reveal the optimal time spent per repetition of the measurement and the loss in precision upon detuning from the optimal reinitialization rate.
(c) Estimate of the decay rate over number of iteratively recorded samples together with estimated uncertainty in orange. Light blue and dashed grey lines show the uncertainty under uniform sampling and optimal sampling according to \cref{eq:exp_optimal_precision}, respectively. 
(d) Logarithmic plot of the estimated uncertainty estimates from subfigure (c). 
The light green line indicates the uncerainty obtained from iterative sampling assuming perfect knowledge of $\true{\kappa}$ and skipping the parameter update step.
(e) Plot of the signal curve and the final samples used to generate the information in subfigures (c) and (d). 
(f) Histogram distribution of the sampling times for samples in subfigure (e). 
The spread partly originates from inaccurate rate estimates as shown in (c).}
\label{fig:exp_iterative}
\end{figure*}

For the present scenario of a single unknown parameter and a single basis function, the determinant of the Hessian reduces to its only entry.
To investigate the behavior of the iterative sampling strategy, perform a numerical experiments with noisy data sampled from a normal distribution.
For simplicity, the signal parameter is treated as dimensionless quantity with value $\true{\kappa} = 1$ such that all sampling times, likewise, are assumed dimensionless. 
The signal-to-noise ratio is chosen as $\true{a}/\sigma = 5$, whereas the exact value barely impacts the qualitative results as long as it is well over \num{1}.
Using two samples taken at $t=0$ and $t=1$ as starting point for the iterative procedure, the estimated decay rate is plotted over the number of samples in \cref{fig:exp_iterative}(c), indicating convergence to the true value with increasing number of samples.
The orange shaded region indicates the estimated uncertainty obtained from the reciprocal Hessian evaluated with \cref{eq:leadingorder_hessian} using the best guess for $\kappa$ to compute $\mmP$ and $\Pperp$.
The blue lines further indicate the expected uncertainty upon taking the same number of samples but distributing them uniformly on the interval $[0,2/\true{\kappa}]$, as indicated in subplot (e) of \cref{fig:exp_iterative}.
While the curves in for the uncertainty estimates shown in \cref{fig:exp_iterative}(d) exhibit the same scaling behavior, the iterative strategy achieves smaller uncertainty over the whole range of number of samples.
With increasing accuracy of the estimate, the estimated uncertainty eventually approaches the light green line which is computed assuming perfect knowledge of the true value $\true{\kappa}$. 
Relative to the uniform sampling, the uncertainty is reduced of up to \qty{25}{\percent} and, thus, aligns with the previous theoretical considerations.
\Cref{fig:exp_iterative}(e) and (f) show the samples generated during the iterative sampling procedure used to generate graphs (c) and (d), where the iterative procedure alternatingly takes samples at $t=0$ and $t>0$ with a ratio of approximately \qty{21}{\percent}, closely resembling the optimal ratio $r_\mathrm{opt}$.
The finite spread for samples at $t>0$ partly stems from inaccuracy in the estimate of $\kappa$, but also is a feature of the optimization with respect to gain in information entropy as shown in~\cref{fig:inf_ent_iteration}(a).

Upon taking the ratio $\true{a}/\sigma$ on the order of 1 or smaller generally is problematic for parameter estimation from the posterior distribution unless one employs a strong prior distribution.
With increasing number of samples, such that the estimate of $\kappa$ is accurate up to approximately \qty{50}{\percent}, above observations about the scaling of uncertainty are recovered for both, the iterative and uniform sampling approach.

\subsection*{Example B: Exponentially Damped Sinusoids}
For an exponentially decaying signal with a single frequency as shown in \cref{fig:demo}(a), a possible choice for the basis functions and corresponding derivatives are given in \cref{tab:osc_basis}.
The true vector $\mva$ to describe the linear parameters, therefore, is of dimension 2 and may be parametrized as $\mva = a_0 \transpose{(\cos(\alpha ), \sin(\alpha ))} $ with $\alpha$ denoting some initial signal phase.

\begin{table}
\caption{Basis functions for an oscillating signal with a single frequency $\omega$ subject to exponential decay with rate $\kappa$ and the respective derivatives.}
\centering
\begin{tabular}{c|c|c}
$\varphi$ & $\partial_\omega \varphi$ & $\partial_\kappa \varphi$ \\
\hline 
$\cos(\omega t) \eulernum^{-\kappa t}$ & $-t \sin(\omega t) \eulernum^{-\kappa t}$ & $-t \cos(\omega t) \eulernum^{-\kappa t}$ \\
$\sin(\omega t) \eulernum^{-\kappa t}$ & $t \cos(\omega t) \eulernum^{-\kappa t}$ & $-t \sin(\omega t) \eulernum^{-\kappa t}$ 
\end{tabular}
\label{tab:osc_basis}
\end{table}

\subsubsection*{Uniform Sampling}
The derivation of approximate expression for the estimation uncertainty for frequency and decay rate of an oscillating signal can be simplified upon making the transition to natural units of time.
Explicitly, the parameters $\omega$ and $\kappa$ are rescaled to absorb the sampling interval $\Delta t$ such that $t$ in the basis functions takes the value $0,\dots,n-1$.
Starting from \cref{eq:Hdiag_simplified} to compute the elements of the leading order Hessian then leads to 
\begin{align}
\begin{split}
\transpose\mmP{} &\mmP =  \\
&\sum\limits_{t=0}^{n-1}\begin{pmatrix}
 \cos^2(\omega t) \eulernum^{-2\kappa t} & \cos(\omega t) \sin(\omega t) \eulernum^{-2\kappa t} \\
\cos(\omega t) \sin(\omega t) \eulernum^{-2\kappa t} & \sin^2(\omega t) \eulernum^{-2\kappa t}
\end{pmatrix}
\end{split} \label{eq:phiphi_osc}
\end{align}
For any reasonable experimental setup that allows to unequivocally identify an oscillating signal as shown in \cref{fig:demo}(a), we would like the recorded data to cover at least one full oscillation ($\true{\omega}n \gtrsim 1$) which ideally concludes within the signal lifetime ($\true{\omega} > 2\pi\true{\kappa}$).
The restriction to above defined regime permits to simplify the expression in \cref{eq:phiphi_osc} by neglecting the fast oscillating terms in the summation as
\begin{align}
\begin{aligned}
\sum\limits_t\cos^2(\omega t) \eulernum^{-2\kappa t} &= \frac{1}{2} \sum\limits_t (1-\cos(2\omega t)) \eulernum^{-2 \kappa t} \\
&\approx \frac{1}{2} \sum\limits_t \eulernum^{-2 \kappa t} \fullstop
\end{aligned} \label{eq:central_approx}
\end{align}
Under the same approximation applied to the remaining matrix elements, the off-diagonal elements are neglected entirely and the inverse of \cref{eq:phiphi_osc} approximately yields
\begin{align}
\big( \transpose\mmP{} \cdot &\mmP \big)^{-1} \approx  \frac{2}{\sum\limits_t \eulernum^{-2 \kappa t}} \idmat \fullstop
\end{align}
With above matrix being diagonal, the evaluation of the matrix elements of $\hess_0$ as in \cref{eq:Hdiag_simplified} reduces to the pair-wise evaluation of the scalar products of $\transpose\mmP{} \partial_j \mmP \mva$.
Invoking the same approximations as in \cref{eq:central_approx}, the phase-dependence of the Hessian cancels naturally and both of the diagonal elements read
\begin{align}
\begin{aligned}
\big[ \hess_o \big]_{\kappa,\kappa} \approx \big[ &\hess_o \big]_{\omega,\omega} \\
&\approx \frac{a_0^2}{\sigma^2} \frac{1}{2} \Big[\sum\limits_{t=0}^{n-1} t^2 \eulernum^{-2 \kappa t} - \frac{\Big(\sum\limits_{t=0}^{n-1} t \eulernum^{-2 \kappa t}\Big)^2}{\sum\limits_{t=0}^{n-1} \eulernum^{-2 \kappa t}} \Big] \fullstop \label{eq:osc_hessian}
\end{aligned}
\end{align}
Off-diagonal terms $\big[ \hess_o \big]_{\kappa,\omega}$ are computed from the fast oscillating terms and can, therefore, in leading order be neglected.

The diagonal terms in $\hess_0$ turn out identical to $1/2$ times the Hessian element for the exponential decay signal under uniform sampling specified in \cref{eq:expmodel_equi}.
The reciprocal of \cref{eq:expmodel_equi} times $1/2$ can be shown to coincide with the the Cramér-Rao lower bound for estimator variance in eq.~(34) of~\cite{Yao1995a}, confirming its validity.
Using the posterior volume measure $\det(\hess_0)$ as figure of merit, the discussions following \cref{eq:expmodel_equi} apply and the optimal total sampling time corresponds to approximately two times the signal lifetime for all possible values of $\true{\omega}$ as long as two or more full oscillations are observed.
The optimal repetition time is indicated for illustrative purposes in \cref{fig:demo}(a) by the gray shaded areas. 
This result, again, is in agreement with statements derived from the Fisher information in context of multichannel impulse responses~\cite{Chardon2010a}.

\subsubsection*{Non-Uniform Sampling and the Iterative Approach}
Because of the periodic nature of oscillating signals, the exact distribution of the samples for selected configurations greatly impacts the estimation performance.
In the most severe case, all samples coincide with roots of the signal and the observer hardly observes any signal variation at all.
The matrix $\transpose\mmP{} \mmP$ becomes near singular and the precision quantified by the Hessian evaluates to values barely larger than zero.
On the other hand, small variation around the signal roots typically exhibit the greatest change in the samples because of the gradient at these locations being maximal.
The discussions following \cref{eq:hess_single_par01,eq:hess_single_par02}, thus, hint at samples at these locations playing an important role for the optimal sampling.
Therefore, the optimal sample distribution is highly nontrivial even for known frequency and phase and general statements are hardly possible.
While estimation performance for stationary oscillating signals always benefits from covering a larger overall time interval~\cite{Bretthorst1988a}, sampling of exponentially decaying signals is expected to exhibit a sweet spot, again.
Using the determinant of the Hessian matrix to quantify the expected performance and recalling the formula for the determinant of a $2\times 2$ matrix, the optimization is expected to yield a sampling scheme that maximizes the precision of individual parameters while keeping the correlations minimal.

\begin{figure*}[!tb]
\centering
\includegraphics[width=\textwidth]{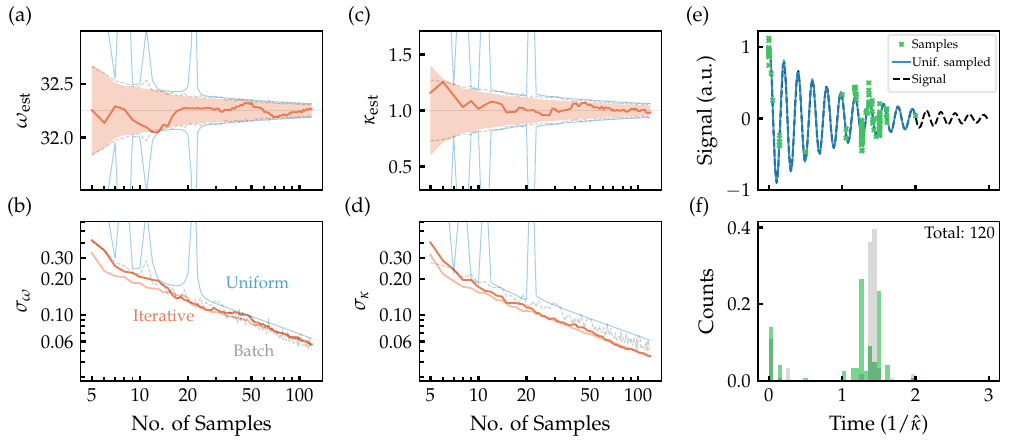}
\caption{(a) Frequency estimate over the number of samples for iterative sampling with the shaded area indicating estimated uncertainty. Blue lines indicate the expected uncertainty for uniform sampling on $t \in [0,2/\true{\kappa}]$, uncertainty obtained from batch optimization is indicated by the gray dashed lines. Divergence in blue curves occur when the sampling frequency is close to the signal frequency and the signal appears non-oscillating to the observer. (b) The uncertainty from (a) (dark orange, blue, and gray) and the expected uncertainty assuming the true parameters perfectly known (light orange). Subfigures (c) and (d) show the corresponding results for the decay rate $\kappa$. (e) Plot of the signal curve and the samples leading to the final estimates. (f) Histogram of the samples in (e) together with the distribution obtained from batch optimization in gray.}
\label{fig:osc_iterative}
\end{figure*}

\Cref{fig:osc_iterative}(a)-(d) shows the parameter estimates for frequency and decay rate as well as corresponding uncertainties over the number of samples.
Over the whole range of considered samples, the uncertainty from the iterative sampling approach is lower than from uniform sampling over the $t\in (0,2/\true{\kappa})$ indicated by the blue lines.
The poles in the blue lines emerge from cases when the sampling frequency approximately coincides with the signal frequency and the underlying signal curve appears constant.
The data points used for the final estimates are plotted over the signal curve in \cref{fig:osc_iterative}(e) and shown in the histogram in \cref{fig:osc_iterative}(f).
With increasing number of samples, the sampling times appear to focus on three points: the origin, a signal maximum close to $1.3/\true{\kappa}$ and a signal root likewise close to $1.3/\true{\kappa}$. 
Similar to the exponential signal scenario, when time points of large amplitude and large gradient fail to coincide, the iterative approach starts alternating the sampling in this scenario with rates optimal to the specific problem.
This indicates that the theoretical optimal sampling strategy also in the oscillating signal scenario potentially also focus on a small number of individual sampling times~\cite{Bolzonello2024a}.

\subsubsection*{Multiple frequencies}
The approach presented in this work can readily be generalized to multiple different frequencies and decay rates, albeit quickly increasing the complexity in the general case. 
The literature concluded that for large separation of individual frequencies the basis functions are approximately orthogonal and the estimation procedure separates~\cite{Bretthorst1988a}.
This argument readily applies also to exponentially decaying signals as long as the basis functions are sufficiently orthogonal taking into account the exponential decay~\cite{Yao1995a}.
For uniform sampling, this requirement is safely achieved if the two signal components achieve a phase shift of at least $\pi$ before either of the signals vanish, compactly expressed through
\begin{align*}
\vert \true{\omega}_1-\true{\omega}_2 \vert / \max(\true{\kappa}_1,\true{\kappa}_2) \gtrsim \pi \fullstop
\end{align*}
Above condition is practically satisfied as soon as the resonances can clearly be identified as separate frequencies after discrete Fourier transformation.
Using the posterior volume measure for precision quantization, the iterative procedure suppresses correlation between different parameters and, therefore, is expected to optimize the orthogonality under discrete sampling.
The exact strategy for this procedure is highly non-trivial and may vastly vary for exact parameter values present.
It is not obvious if non-uniform sampling is more efficient in establishing orthogonality in the sense of requiring fewer samples.
Numerical experiments demonstrate that optimized sampling, again, weighs different signal regions differently, but do not reveal any trivial strategy~\cite{Bolzonello2024a}.

\section{Discussion}
Approximative evaluation of the information for two examples demonstrates and quantifies the potential improvement over uniform sampling of \qtyrange{10}{20}{\percent}.
The explicit dependence of the sampling schemes of true underlying signal parameters is addressed through the application of adaptive sampling schemes, achieving improvements over uniform sampling on similar improvements levels.
The overall scaling laws in the reduction of estimation uncertainty with increasing number of samples remain untouched.

Some of the results presented with the examples in \cref{sec:results} coincide with results obtained from optimal experimental design based on the Fisher information matrix~\cite{Wigren1991a,Han2003a,Chardon2010a}, thereby confirming the validity.
The Bayesian approach employed in this work, in contrast, makes use of the Shannon entropy for the quantification of estimation quality and allows for extension by additional prior distributions on the non-linear parameters.
While the principles of adaptive design optimization have been extensively discussed in the literature, to knowledge of the authors there is no explicit example that applies the ansatz to marginal distributions and addresses the consequences.

The schemes presented in this work focus on scenarios in which the cost of each sample is weighted equally.
There are various extensions of the objective function as in \cref{eq:maxdetH0} possible to account for varying cost, e.g., for samples that require longer waiting times until the measurement.
One of these options has been explored in example A, where the noise parameter $\sigma$ has been adjusted to account for repetition of sampling at the same locations.
Compact formulations for individual weighting of samples in the general, however, require further efforts that are outside the scope of this work.

The simplification undertaken in \cref{eq:Fas} under what we called the asymptotic limit represents a strong approximation that renders inaccurate in scenarios of very poor statistics, i.e., number of samples on the order of free model parameters and low SNR, and Markov-Chain Monte-Carlo sampling is the preferred choice for an accurate representation of the posterior distribution.
In scenarios that permit the use of adaptive schemes, however, the acquisition of samples continuously improves the accuracy of the approximation, eventually moving into the asymptotic limit.
For the examples considered in this work, in particular, the approximation turns out simple and effective.

Uniform sampling generally has the advantage of requiring little to zero prior knowledge about the expected signal structure and the obtained data is particularly accessible for interpretation by human researchers.
The strong dependence of optimized sampling schemes require a predefined signal model and some prior knowledge about the expected parameter values, such that they are most effective for routine experimental setups that are repeated various times.
Therefore, the framework and derived schemes in this work are, for example, particularly relevant to calibration procedures and experiments with a high repetition rate of similar experiments for reliable statistics, such as for anomaly detection.

\section*{Acknowledgements}
LHB thanks Koenraad Audenaert for useful discussions on globally optimal schemes and associated proofs.
This work was supported by the BMBF through grant no 03ZU1110FF (QMED) and no 13N16447 (QuE-MRT).
The European Research Council provided support through the ERC Synergy Grant HyperQ (Grant No. 856432) and project C-QuENS (Grant No. 101135359).

\printbibliography[heading=bibintoc] 

\appendix
\section*{Appendix}
\titleformat
	{\section} 
	[hang] 
	{\large\bfseries} 
	{\Alph{section}} 
	{0.5em} 
	{} 
	[]
\renewcommand{\theequation}{\thesection.\arabic{equation}}
\setcounter{equation}{0}
\section{Update of the Asymptotic Posterior Distribution and Parameter Estimates} \label{app:asym_post_update}
In the fundamental Bayesian update scheme assuming statistically independent data, the posterior distribution is updated upon multiplication by the likelihood for new observations.
From the form of the posterior in \cref{eq:log_marg_likelihood} it becomes apparent, that despite statistically independent observations, the posterior does not separate into previous and new samples after marginalization any more.
Formally, the addition of a new data sample is described by the transformation
\begin{align}
\mmP \to \mmP^\prime = \begin{pmatrix} \mmP \\ \transpose\mvp{} \end{pmatrix} \ \ \text{ and } \ 
\mvy \to \mvy^{\prime} = \begin{pmatrix} \mvy \\ \gamma \end{pmatrix} \comma 
\end{align}
respectively, where $\mvec{\varphi}$ describes the basis functions evaluated at the new time point and $\gamma$ is the newly acquired sample.

To compute the impact onto the orthogonal projector $\Pperp^{\prime}$, recall that the projector $\projector^{\prime} $ is computed using the updated $\mmP^{\prime}$ explicitly as
\begin{align}
\projector^{\prime}= \mmPp \big( \mmPtp \mmPp \big)^{-1} \mmPtp \fullstop \label{eq:Pperp_update01} 
\end{align}
Noting that the update to $\transpose\mmP{} \mmP$ can be written as a rank-1 update, application of the Sherman-Morrison formula~\cite{Sherman1949a} yields
\begin{align}
\begin{aligned}
\big( \mmPtp \mmPp \big)^{-1} 
 &= \big( \transpose\mmP{} \mmP + \mvec{\varphi} \transpose\mvec{\varphi} \big)^{-1} \\
&= \big( \transpose\mmP{} \mmP \big)^{-1} - \frac{( \transpose\mmP{} \mmP )^{-1} \mvec{\varphi} \transpose\mvec{\varphi} \big( \transpose\mmP{} \mmP \big)^{-1}}{1 + \transpose\mvec{\varphi} \big( \transpose\mmP{} \mmP \big)^{-1} \mvec{\varphi}} \fullstop
\end{aligned}
\end{align}
Introducing the vector
\begin{align}
\mvu := \mmP (\transpose\mmP{} \mmP)^{-1} \mvec{\varphi}
\end{align}
and the scalar
\begin{align}
\alpha := \transpose\mvec{\varphi} (\transpose\mmP{} \mmP)^{-1} \mvec{\varphi} \comma
\end{align}
the projector from \cref{eq:Pperp_update01} may be rewritten as
\begin{align}
\begin{aligned}
\projector^{\prime} &= \begin{pmatrix}
\projector & \mvu \\
\transpose\mvu{} & \alpha
\end{pmatrix} 
- \frac{1}{1 + \alpha}
\begin{pmatrix}
 \mvu \transpose\mvu{} & \alpha \mvu \\
\alpha \transpose\mvu{} & \alpha^2
\end{pmatrix} \\
&= \begin{pmatrix}
\projector & 0 \\
0 & 0
\end{pmatrix} 
+ \frac{1}{1 + \alpha}
\begin{pmatrix}
- \mvu \transpose\mvu{} & \mvu \\
\transpose\mvu{} & \alpha
\end{pmatrix} \fullstop
\end{aligned} 
\end{align}
The orthogonal projector from $\Pperp^{\prime} = \idmat - \projector^{\prime}$ then likewise separates into previous and new contributions as
\begin{align}
\Pperp^{\prime} = \begin{pmatrix}
\Pperp & 0 \\
0 & 0
\end{pmatrix} 
+ \frac{1}{1 + \alpha}
\begin{pmatrix}
 \mvu \transpose\mvu{} & -\mvu \\
- \transpose\mvu{} & 1
\end{pmatrix} \fullstop \label{eq:Pperp_update}
\end{align}
To compute the value of $F_\mathrm{as} \varpropto \Vert \Pperp^{\prime} \mvy^{\prime} \Vert^2$ after the update, note that the product of the two terms contribution to above $\Pperp^{\prime} $ yields zero, such that the final result separates into a contribution from previous samples and an additional term.
Algebraic simplification and noting that $\mvu = \transpose(\mmP^\dagger) \mvec{\phi}$ ultimately yields
\begin{align}
\Vert \Pperp^{\prime} \mvy^{\prime} \Vert^2 = \Vert \Pperp \mvy \Vert^2 + \frac{\big(\transpose\mvec{\phi} \mmP^\dagger \mvy - \gamma\big)^2}{1+\alpha} \fullstop 
\label{eq:obj_update}
\end{align}
The addition of a new sample extends the previous value by the squared distance between the best guess for the signal value and the sample value $\gamma$ weighted by $1/(1+\alpha)$.
Thus, applied explicitly to the posterior distribution in the asymptotic limit yields
\begin{align}
F_\mathrm{as}(\omega;\mvy^{\prime}) = F_\mathrm{as}(\omega;\mvy) + \Delta(\omega;\mvy^{\prime}) \comma \label{eq:Fas_update}
\end{align}
with $\Delta$ representing the second term of \cref{eq:obj_update} multiplied by $1/2\sigma^2$.
Assuming the posterior distribution to be governed by the exponential of above expression, the distribution separates into a product of the previous distribution and an update term.
Previously eliminated linear parameters $\mva$ take the role of latent variables that relate all individual data samples, such that the update term itself explicitly depends on all previously observed samples and, therefore, does not follow the ideal case for Bayesian updating.
The requirement to keep book of all previous observations to compute the update to the posterior makes plainly recomputing the posterior distribution taking into account all new samples the preferred method in many practical scenarios.

To compute the update to the leading order Hessian, recall from \cref{eq:leadingorder_hessian} that the contribution to the matrix elements of the Hessian at location of the local maximum featuring the sampling points is computed from
\begin{align}
\transpose(\Pperp \mvx_j) \cdot (\Pperp \mvx_k) \comma \label{eq:recall_hessian}
\end{align}
with $\mvx_j = \partial_j\mmP\true{\mva}$. 
The addition of a sample adds $\eta_j = \transpose(\partial_j\mvec{\varphi}) \mva$ as additional component to $\mvx_j$ and transforms $\Pperp$ to $\Pperp^{\prime}$ as in \cref{eq:Pperp_update}.
Now noting that the symmetric projector satisfies $(\Pperp^{\prime})^2 = \Pperp^{\prime}$ and recalling that $\alpha$ can be rewritten as $\alpha=\transpose\mvu{} \mvu$, the insertion of $\Pperp^{\prime}$ into \cref{eq:recall_hessian} leads to the expression in \cref{eq:hess_update} of the main text.

\section{Numeric Stability of the Iterative Sampling} \label{app:num_stab}
\begin{figure}
    \centering
    \includegraphics[scale=1]{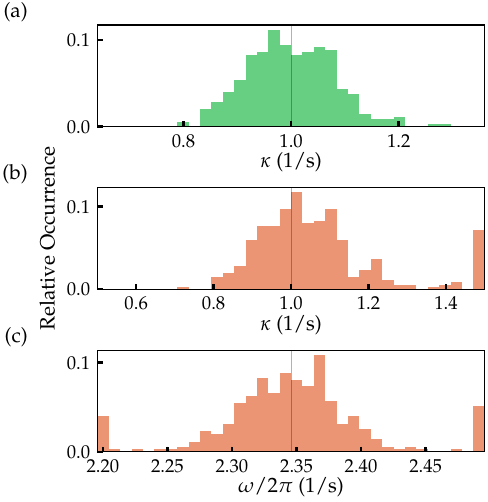}
    \caption{Histograms of parameter estimates for the exponentially decaying signal in (a) and for the oscillating signal in (b)-(c) after \num{100} iteratively obtained samples. Under- and overflow bins make up approximately \qty{7.1}{\percent} and \qty{9.1}{\percent} in histograms (b) and (c), respectively. Vertical lines indicate true values of the respective parameters.}
    \label{fig:succes_plot}
\end{figure}

To investigate the sampling stability of the iterative approach, the procedure is repeated \num{350} times for examples A and B presented in \cref{sec:results} for different realizations of the noise and signal-to-noise ratio of $\Vert \true{\mva} \Vert / \sigma = 5$.
Starting from \num{3} and \num{5} initial samples taken uniformly on the interval $[0,2/\true{\kappa}]$, respectively, parameter estimates after \num{100} samples for example A are plotted in \cref{fig:succes_plot}(a) and for example B in \cref{fig:succes_plot}(b-c).

For the simple exponential decay in example A, the parameter estimates converged to the true parameter values in all of the runs considered.
Occasionally, if the initial samples suggest parameter estimates that deviate strongly from the true value and the optimizer becomes stuck in a local optimum and convergence to the true parameter may never be achieved.
This flaw may be mitigated through the inclusion of strong parameter bounds in the estimation or regularization via informative prior distributions.
Increasing the number of uniformly distributed initial samples may stabilize the fitting and sampling procedure. 
The sampling procedure for example B mostly suffered from aliasing effects under which the estimates appear to converge to reasonable values that are multiples of the actual frequency, leading to an overall success rate of approximately \qty{90.9}{\percent}, as indicated by the overflow bins in \cref{fig:succes_plot}(c) constituting approximately \qty{9.1}{\percent} of the runs.
The effects can be reduced by the same measures as before: stronger parameter bounds imposed on the fitting procedure, regularization through use of the appropriate prior distributions or increase of the number of initial samples.
The fail rate is expected to grow with the number of free non-linear parameters in the estimation problem and can, eventually, only be mitigated through strong priors or a sufficient amount of initial samples.
Ultimately, for the iterative procedure to render useful, the number of initial samples and expected number of total samples to satisfy the stopping criterion have to be balanced reasonably.

\section{Optimal Sampling Configuration for the Exponentially Decaying Signal} \label{app:optsamp_expdecay}
The optimal sampling configuration for an exponentially decaying signal is provided with the global maximum of $\phi$ with 
\begin{align}
\phi( \lbrace t_j \rbrace) = S_2 - S_1^2 / S_0 \label{eq:phidef}
\end{align}
where $S_l = \sum_{j=1}^{n} t_j^{l} \eulernum^{-2\kappa t_j}$ and $t_j\geq 0$ for $j=1,\dots,n$.
Let the function $\phi$ be as in \cref{eq:phidef}. 
The maximum of $\phi$ is achieved with each $t_j$ assuming either of $t=0$ and $t=t^\ast$, where $t^\ast$ is implicitly defined via
\begin{align}
t^{\ast} = \frac{1}{\kappa} + \frac{S_1}{S_0} \fullstop
\end{align}
Proof of this statement follows below.

Note that $\phi$ is continuous in each variable $t_j \geq 0$ and bounded for finite $n$.
To argue that $\phi$ exhibits a global maximum on a compact set $\mathcal{T} \subset \mathbb{R}_+^n$, fix all $t_j$ but the one with index $l$ and show that there exists a point $0 \leq t_l<\infty$ for which $\phi$ assume a value larger than for taking $t_l \to \infty$.  
Separating out contributions from $t_l$, $\phi$ can be rewritten as
\begin{align}
    \phi = (\tilde{S}_2 + t_l^2 \eulernum^{-2\kappa t_l}) - \frac{(\tilde{S}_1 + t_l \eulernum^{-2\kappa t_l})^2}{\tilde{S}_0 + \eulernum^{-2\kappa t_l}} \comma
\end{align}
where the tilde indicates quantities evaluated using only all remaining $n-1$ parameters $t_j$.
Using above expression, one finds $\phi(t_l=0) - \phi(t_l \to \infty) = \tilde{S}_1^2/\tilde{S}_0(\tilde{S}_0+1)$, which is larger or equal to \num{0} and assumes \num{0} exactly only when all remaining $t_j$ also assume the value \num{0}. 
Because $\phi$ is invariant under permutation of $t_j$, for each $t_j$ there exists at least one point that is larger than the bound for taking the corresponding $t_j$ to $\infty$ and, therefore, $\phi$ assumes its global maximum on a compact subset of $\mathbb{R}_+^n$. 
Therefore, there exists a global maximizer of $\phi$ on the feasible set $\lbrace t_j \geq 0 \rbrace$.

\begin{figure}
\centering
\includegraphics[scale=1]{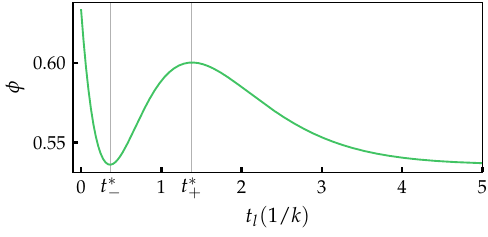}
\caption{Schematic plot of $\phi( \lbrace t_j \rbrace ) $ after fixing all $t_j$ but one, illustrating that $\phi$ always takes higher values at $\tp$ compared to $\tm$.}
\label{fig:phiplot}
\end{figure}

To find the explicit global maximizer, note that the configuration of all $t_j=0$ yields $\phi=0$ which is not the maximum and, therefore, this point is excluded from further considerations.
Assume that the maximizer features some parameters on the boundary of the interval $t_j = 0$ and some inside $t_j>0$.
A sufficient condition for a valid maximizer with respect to the boundary points constitutes the derivative of $\phi$ with respect to corresponding $t_j$ evaluated at the maximizer must be smaller than zero, i.e., moving these $t_j$ into the interior of the feasible set decreases $\phi$.
For interior points, in contrast, the derivative of $\phi$ must evaluate to zero to indicate a local extremum.
Computing the derivative of $\phi$ with respect to a particular $t_l$ can be simplified to
\begin{align}
\begin{aligned}
\frac{\partial \phi}{\partial t_l} = - \frac{2 \eulernum^{-2 \kappa t_l}}{S_0^2} \big\lbrace  S_0^2 \kappa t_l^2 - &(S_0^2 + 2S_0S_1 \kappa) t_l \\
&+ (S_0 S_1 + \kappa S_1^2) \big\rbrace \fullstop \label{eq:derphi}
\end{aligned}
\end{align}
Taking into account that $S_0,S_1,S_2 > 0$ for any configuration of $t_j$ other than all $t_j=0$, one can immediately read off that \cref{eq:derphi} assumes a negative value for $t_l = 0$ and the condition for arguments on the boundary of the feasible set is always satisfied in this case.
To find the critical interior points from $\partial \phi / \partial t_l = 0$, one can solve the quadratic equation in parenthesis of \cref{eq:derphi} and obtain two implicit solutions for $t_l$ as
\begin{align}
t^{\ast}_{-} = \frac{S_1}{S_0} \ \ \wedge \ \ t^{\ast}_{+} = \frac{1}{\kappa} + \frac{S_1}{S_0} \fullstop
\end{align}
This already collapses the space of possible maximizers to be composed of three different values for each $t_j$.
To further refine the solution, fix all $t_j$ but one to \num{0} or either of $t^{\ast}_\pm$ and employ $\tilde{S}_0,\tilde{S}_1,\tilde{S}_2$ for the fixed $t_j$, again. 
The critical values of the remaining parameter are implicitly defined through
\begin{align}
t_-^{\ast} = \frac{\tilde{S}_1 + t_-^{\ast} \eulernum^{-2 \kappa t_-^{\ast}}}{\tilde{S}_0 + \eulernum^{-2\kappa t_-^{\ast}}} \comma \label{eq:tmimpl}
\end{align}
which is satisfied by $t_-^{\ast} = \tilde{S}_1 / \tilde{S}_0$.
Similarly, $t_+^{\ast}$ is defined by
\begin{align}
t_+^{\ast} =  \frac{1}{\kappa} + \frac{\tilde{S}_1 + t_+^{\ast} \eulernum^{-2 \kappa t_+^{\ast}}}{\tilde{S}_0 + \eulernum^{-2\kappa t_+^{\ast}}} \comma \label{eq:tpimpl}
\end{align}
which is satisfied for $t_+^{\ast} = 1/\kappa + \tilde{S}_1 / \tilde{S}_0 + \delta$, with the exact value of $\delta > 0$ being irrelevant to the proof.
Now use these values to evaluate
\begin{align}
\begin{split}
\phi(\tp ) - \phi(\tm) &= \tilde{S}_2 + (\tp)^2 \eulernum^{-2k\tp} - \frac{(\tilde{S}_1 + \tp \eulernum^{-2k\tp})^2}{\tilde{S}_0 + \eulernum^{-2k\tp}} \\
&- \big\lbrace \tilde{S}_2 + (\tm)^2 \eulernum^{-2k\tm} - \frac{(\tilde{S}_1 + \tm \eulernum^{-2k\tm})^2}{\tilde{S}_0 + \eulernum^{-2k\tm}} \big\rbrace \fullstop
\end{split}
\end{align}
Inserting the values for $\tm$ and $\tp$ derived from \cref{eq:tmimpl,eq:tpimpl}, respectively, simplifies the difference to
\begin{align}
\frac{\tilde{S}_0 (1+\kappa \delta)^2}{\kappa^2 ( 1+\tilde{S}_0 \exp(2+2\kappa(\delta+\tilde{S}_1/\tilde{S}_0)))} \fullstop
\end{align}
Above expression is evidently positive, indicating that $\phi(\tp ) > \phi(\tm)$ upon fixing all remaining $t_j$.
This also implies, that $\phi(\lbrace t_j \rbrace)$ always increases when moving each individual $t_j$ that previously assumed value $\tm$ to $\tp$.
Therefore, the maximizer is composed of $t_j$ assuming values $0$ or $\tp$, completing the proof.
Because of the invariance of $\phi$ under permutation of the $t_j$, the actual maximum is heavily degenerate.
\Cref{fig:phiplot} illustrates the behavior of $\phi$ when fixing all values of $t_j$ but one.

\end{document}